\documentclass[12pt,onecolumn,amsmath,amssymb, spacing=1.5pt]{article}
\usepackage[letterpaper,margin=1in]{geometry}
\usepackage{ulem}

\newcommand{\br}{\boldsymbol{\textbf{r}}}

\newcommand{\bL}{\boldsymbol{\textbf{L}}}

\newcommand{\bR}{\boldsymbol{\textbf{R}}}
\newcommand{\bS}{\boldsymbol{\textbf{S}}}

\newcommand{\rhoin}{\rho_{\mathrm{in}}}
\newcommand{\rhoout}{\rho_{\mathrm{out}}}

\newcommand{\dr}{\,d\br}

\newcommand{\rhod}{\rho_{\mathrm{data}}}

\newcommand{\vKS}{v_{\mathrm{KS}}}
\newcommand{\vH}{v_{\rm{H}}}

\newcommand{\vext}{v_{\rm{ext}}}
\newcommand{\vxc}{v_{\rm{xc}}}
\newcommand{\vxcalpha}{v_{\mathrm{xc}, \alpha}^{\mathrm{rem}}}

\newcommand{\Exc}{E_{\rm{xc}}}

\newcommand{\EZZ}{E_{\mathrm{zz}}}

\newcommand{\Ees}{E_{\mathrm{es}}}
\newcommand{\EH}{E_{\mathrm{H}}}

\newcommand{\Exx}{E_{\mathrm{xx}}}
\newcommand{\Ts}{T_{\mathrm{s}}}
\newcommand{\EGS}{E_\mathrm{GS}}
\newcommand{\psiini}{\psi_i^{\mathrm{in}}}
\newcommand{\psiouti}{\psi_i^{\mathrm{out}}}
\newcommand{\VXACE}{\hat{V}_{\mathrm{X}}^{\mathrm{ACE}}}

\newcommand{\HGKS}{\hat{H}_{\mathrm{GKS}}}
\newcommand{\VX}{\hat{V}_{\mathrm{X}}}
\newcommand{\VXin}{\hat{V}_{\mathrm{X}}^{\mathrm{in}}}
\newcommand{\Excalpha}{E_{\mathrm{xc},\alpha}^{\mathrm{rem}}}
\newcommand{\vloca}{v_{\alpha}^{\mathrm{loc}}}
\newcommand{\EMF}{E_{\mathrm{MF}}}

\newcommand{\invDFT}{\texttt{invDFT}~}

\usepackage{doi}
\usepackage{mathtools}
\usepackage{footnote}
\usepackage{hyperref}
\hypersetup{
    colorlinks=true,
    linkcolor=magenta,
    filecolor=magenta,      
    urlcolor=cyan,
    }

\usepackage{subfiles}
\usepackage{graphicx}
\usepackage{graphics}
\usepackage{multirow}
\usepackage{amsxtra}
\usepackage{stmaryrd}
\usepackage{mathrsfs}
\usepackage{caption}
\usepackage{subcaption}
\usepackage{epsfig}
\usepackage{rotating}
\usepackage{setspace}
\usepackage{float}
\usepackage{amsfonts}
\usepackage{amsbsy}
\usepackage{amscd}
\usepackage{amsthm}
\usepackage{relsize}
\usepackage{color}
\usepackage{tablefootnote}
\usepackage{tabularx}
\usepackage{caption}
\usepackage{sidecap}
\usepackage{dcolumn}
\usepackage{algorithm}
\usepackage{algorithmic}
\usepackage{physics}
\usepackage{breqn}
\usepackage{lineno}

\usepackage{authblk}

\definecolor{hellgruen}{rgb}{0.2,0.7,0.2}

\newcolumntype{M}[1]{>{\centering\arraybackslash}m{#1}}
\newcolumntype{N}{@{}m{0pt}@{}}

\begin{document}

\title{Exchange-Correlation Potentials and Energies from Inverse Generalized Kohn-Sham Calculations}


\author[a,b]{Vishal Subramanian}
\author[a] {Bikash Kanungo}
\author[c] {Vaibhav Khanna}
\author[c]{Paul M. Zimmerman}
\author[a,d, *]{Vikram Gavini}
\affil[a]{\small Department of Mechanical Engineering, University of Michigan, Ann Arbor, Michigan 48109, USA}
\affil[b]{\small MICDE, University of Michigan, Ann Arbor, Michigan 48109, USA}
\affil[c]{\small Department of Chemistry, University of Michigan, Ann Arbor, Michigan 48109, USA}
\affil[d]{\small Department of Materials Science and Engineering, University of Michigan, Ann Arbor, Michigan 48109, USA}
\affil[$*$]{Corresponding author: vikramg@umich.edu}
\date{}

\maketitle


\begin{abstract} 
The Kohn-Sham (KS) formulation of density functional theory (DFT) is a map from the many-electron problem to an effective single-electron problem that is governed by a local multiplicative potential. The generalized-Kohn-Sham (GKS) formalism extends it to permit any single-electron operator---nonlocal, local non-multiplicative, local multiplicative, or any combination of them. Doing so expands the scope and ease of modeling the exchange-correlation (XC) functional in DFT, which encodes the complicated many-electron interactions into a mean-field of the electron density. However, unlike KS theory, development of XC functionals in GKS theory has been hindered by the absence of corresponding exact XC potentials and energies. We present the exact XC potentials and energies for atoms and molecules by solving the inverse GKS problem, using highly accurate correlated \textit{ab initio} densities. Our approach is validated across weakly and strongly correlated systems. We further examine a common, yet untested, assumption that KS and GKS correlation potentials and energies are similar, finding instead that they differ substantially in strongly correlated systems. Overall, this work offers a powerful tool to model next-generation of XC functionals within the GKS formalism of DFT.
\end{abstract}


\section{Introduction}\label{intro}

Density functional theory (DFT) is one of the most consequential development in electronic structure theory. It remains the most popular electronic structure method, predicting a range of properties of atoms, molecules, and solids~\cite{becke2014perspective}.  DFT establishes all groundstate properties of materials as a universal functional of the groundstate electron density ($\rho(\br)$) ~\cite{PhysRev.136.B864}. Practically, its prowess is realized through the construction of an auxiliary system of non-interacting electrons, called the Kohn-Sham (KS) system, which yields the same groundstate density as the interacting system ~\cite{kohn1965self}. The KS formalism reduces the interacting many-electron problem to an effective single electron problem, where the electrons interact through a mean-field potential defined in terms of the electron density ($\rho(\br)$) and termed as the Kohn-Sham potential ($\vKS[\rho](\br)$).  The most vital ingredient in $\vKS$ is the exchange-correlation (XC) potential ($\vxc$), which encapsulates the quantum many-body interaction between the electrons as a universal functional of the density ($\rho$). The exact form of $\vxc$ is unknown, needing approximations. Although many useful approximations have been proposed over the past 50 years, DFT still remains far from the chemical accuracy attained through quantum many-body methods, such as configuration interaction (CI) ~\cite{shavitt1977method}, coupled-cluster (CC)~\cite{bartlett2007coupled}, or quantum Monte Carlo (QMC) ~\cite{foulkes2001quantum}. Thus, the all-important challenge in DFT is to model better XC functionals.

\noindent The quest for better XC has led to a new auxiliary system, called the generalized Kohn-Sham (GKS) system\cite{seidl1996generalized}. Similar to the KS system, the the GKS system is still an effective single electron system, and hence, defined via a single or ensemble of degenerate Slater determinants, termed as the GKS Slater determinant. However, unlike the local multiplicative nature of the mean-field potential ($\vKS$) in the KS formalism, the GKS formalism allows for any single electron operator defined in terms of the GKS Slater determinant. The prototypical GKS system is defined by the ``hybrid" XC functional (e.g., B3LYP~\cite{Becke1993a,Becke1993b,Lee1988}, PBE0~\cite{Adamo1999}), where the single electron operator involves a fraction of the nonlocal exact (Hartree-Fock) exchange operator and a local multiplicative potential. A more recent GKS system is defined by the meta-GGA XC functional (e.g., SCAN~\cite{Sun2015}, r2SCAN~\cite{Furness2020accurate}, etc.), where the single electron operator includes a combination of local muliplicative potential and a local non-multiplicative operator, such as the derivative operator. It is easy to see that the KS system is a special case of the GKS formalism.  The GKS formalism enjoys both conceptual and practical advantages over the KS  formalism. First, GKS, by construction, provides greater flexibility in designing XC functionals. Second, the ability to explicitly use the KS Slater determinant allows to embed a large portion of known electronic interactions, and hence, ease the modeling of the remainder XC functional. These advantages of the GKS formalism reflects in their superior accuracies in a wide range of thermochemical~\cite{Goerigk2017look} and solid-state~\cite{Zhang2018performance} properties.

\noindent Despite the success of GKS formalism, fundamental deficiencies, such as self-interaction and static correlation errors still persist~\cite{Cohen2012}. As a result, XC functionals that offer chemical accuracy for both weakly and strongly correlated systems still remain elusive. To that end, the \textit{inverse} DFT problem~\cite{gorling1992, wang1993, zhao1994electron, Leeuwen1994, Tozer1996, wu2003direct, Jacob2011, gould2014, Ryabinkin2015, Jensen2018, kanungo2019exact, Kumar2019,  Shi2021, Erhard2022, tribedi2023exchange,  gould2023, Aouina2023, khanna2025exchange, Visagan2025} of determining the exact XC potential ($\vxc$) corresponding to a given electron density ($\rho(\br)$), typically obtained from accurate quantum many-body methods (e.g., CI, CC, or QMC), provides a powerful framework for studying the nature of the XC functional~\cite{Khanna2026bridges}. Recently, the inverse DFT problem has also been instrumental in providing high-fidelity data to machine-learn accurate XC functionals~\cite{Schmidt2019machine, Kanungo2025learning}. However, almost all attempts at inverse DFT are done within the KS formalism. The only attempt at inverse DFT within GKS formalism~\cite{garrick2020exact} is limited to single atoms and ions, where the spherically symmetry reduces the 3D problem to 1D radial problem. Thus, a generic 3D inverse GKS method that is applicable to polyatomic systems is lacking. This work seeks to fill that gap.

\noindent Inverse GKS poses several numerical and computational challenges. Computationally, inverse GKS is far more challenging than inverse KS, as it demands the evaluation of the nonlocal exchange operator. Numerically, it suffers from three sources of ill-posedness, all of which are inherited from the much simpler inverse KS problem. The ill-posedness, in turn, leads to unphysical and/or non-unique potentials. The first source of ill-posedness stems from the finite (incomplete) nature of the atomic orbital (AO) basis (e.g., Gaussian or Slater basis). The finite basis can lead to artificial small eigenvalues in the discrete GKS linear response function $\chi_{\mathrm{GKS}}(\br,\br')=\frac{\delta \rho(\br)}{\delta \vxc(\br')}$, which measures the change in the density due to a perturbation in XC potential. In other words, finite basis permit certain highly oscillatory perturbations to the $\vxc$ which leaves the density unchanged. The second source of ill-posedness lies in the, often, ``unbalanced" nature of the orbital and potential basis. Given a basis $\{N_i\}$ for the GKS orbitals, all $\vxc$'s for which the matrix $V_{ij}=\int N_i(\br) \vxc(\br) N_j(\br)\dr$ are the same yields the same density. This ill-posedness can, to a large extent, be alleviated by using a ``balanced" potential basis: a potential basis that has no component orthogonal to the space spanned by the product of the orbital basis (i.e., space spanned by $N_i N_j$). The third source of inaccuracy stems from the the target density itself. The target density from quantum many-body methods (e.g., CI, CC), being represented in a Gaussian or Slater basis, do not accurately define the densities near the nuclei and can exhibit incorrect decay (e.g., Gaussian decay instead of exponential decay when using Gaussian basis). These incorrect asymptotics often induce spurious oscillations in the resulting XC potentials.

\begin{figure}[htbp!]
    \centering
    \includegraphics[scale=1]{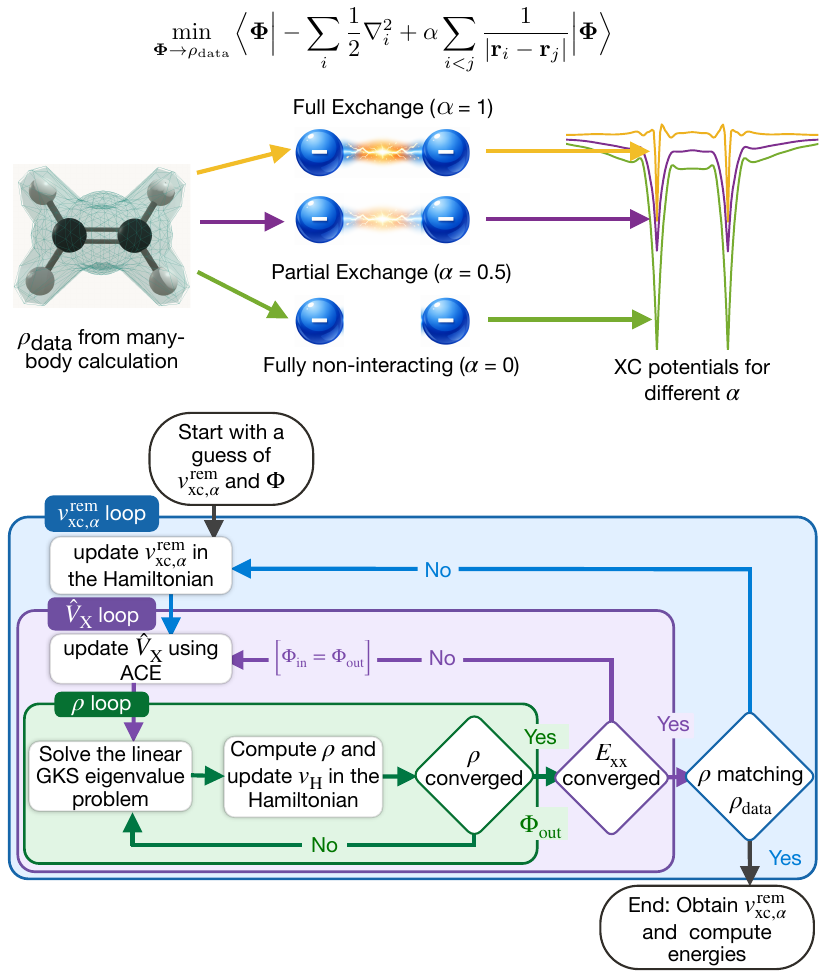}
    \caption{Top panel: Schematic of the inverse GKS problem. Here the inversion is restricted to the prototypical GKS system containing a fraction ($\alpha$) of the exact exchange operator. Bottom panel: A schematic of the nested loops in the inverse GKS. }
    \label{fig:gks_schema} 
\end{figure}

\noindent We address the above numerical and computational challenges through a combination of finite-element (FE) basis and adaptively compressed exchange (ACE) method. The FE basis, being systematically convergent, makes the discrete GKS linear response ($\chi_{\mathrm{GKS}}$) closer to the continuous, and hence, avoids any artificial small eigenvalues. The FE basis also allows for ``balanced" potential basis owing to its flexibility of the polynomial order of the basis. To elaborate, we use higher-order (polynomial order 4-5) finite-elements for the orbital basis and linear finite-elements (polynomial order 1) for the potential basis. This ensures that the no component of the potential basis is orthogonal to the product of the orbital basis. Lastly, as will be discussed later, the FE basis can be used to also resolve artifacts arising from incorrect asymptotic behavior in the target density~\cite{kanungo2019exact, Kanungo2021comparison, Kanungo2023}. 
The ACE method~\cite{lin2016adaptively} allows for efficient evaluation of the action of the nonlocal exchange operator on a set of orbitals, thereby making inverse GKS calculations computationally feasible. We demonstrate the robustness and accuracy of our proposed approach on a range both weakly and strongly correlated molecules. These advances allow us to test a conventional yet, largely untested, assumption that the KS and GKS XC potentials and mean-field energies (sum of non-interacting kinetic, electrostatic, and exchange energies) are similar to one another. We find that, while this assumption holds for weakly correlated systems, it breaks down significantly for strongly correlated systems (e.g., stretched molecules). All these advances for inverse GKS are implemented within a development branch of the massively parallel CPU-GPU \invDFT software package~\cite{subramanian2026invdft}. We refer to Fig. \ref{fig:gks_schema} for an overview of the work. We envisage the combination of the exact XC potentials, exact XC energies, and the GKS orbitals resulting from our approach to serve as high fidelity training data to machine-learn accurate XC functional that can significantly
boost the predictive powers of DFT in chemical, physical, and materials sciences.

\section{Results} \label{sec:results}
\subsection*{Inverse GKS Formalism}
Given a target groundstate density $\rhod(\br)$ and a prescribed nonlocal or local non-multiplicative operator, the inverse GKS problem is to determine the local multiplicative XC potential that yields $\rhod(\br)$. For simplicity, we restrict to the prototypical GKS system containing a fraction ($\alpha$) of the exact exchange operator, in addition to the local multiplicative XC potential we seek. Here, we present the formulation for a closed-shell (spin-unpolarised) and  finite systems (atoms and molecules) which have real-valued orbitals. Extension to other forms of nonlocal or local non-multiplicative operators, open-shell (spin-polarized), and/or extended (solids) systems can be built upon the proposed method.  The XC energy in such a GKS system is defined as 
\begin{equation} \label{eq:gksInv_Exc}
E_{\mathrm{xc}}^\alpha =
\alpha \Exx[\{\psi_i\}] + \Excalpha[\rhod]\,, \quad \Exx = -  \sum_{i=1}^{N_e/2}\sum_{j=1}^{N_e/2}\int\int\frac{\psi_i(\br)\psi_j(\br)\psi_j(\br')\psi_i(\br')}{|\br-\br'|}\dr\dr'\,,
\end{equation}
where $N_e$ is the number of electrons, $\Exx$ is the exact exchange energy defined in terms of the GKS orbitals $\psi_i$ corresponding to $\rhod$, and $\Excalpha$ denotes the remainder density-dependent XC energy. The inverse GKS seeks to find the remainder local multiplicative XC potential $\vxcalpha(\br)=\frac{\delta \Excalpha[\rhod]}{\delta \rhod(\br)}$ as well as the GKS orbitals $\psi_i$. This can be posed as the following Levy-Lieb constrained optimization problem~\cite{levy1979universal,lieb1983density, seidl1996generalized}: 
\begin{equation} \label{eq:LevyLieb}
\min_{\Phi\rightarrow\rhod} 
    \Big\langle \Phi \Big| - \sum_i \frac{1}{2} \nabla_i ^2  + \alpha  \sum_{i < j} \frac{1}{|\br_i - \br_j |} \Big| \Phi \Big\rangle\,,
\end{equation}
where $\Phi$ is the Slater determinant comprised of the GKS orbitals $\psi_i$ and $\Phi\rightarrow\rhod$ denotes all Slater determinants which yield $\rhod$. The above constrained optimization can be recast as an unconstrained optimization by introducing a local multiplicative potential ($\vloca(\br)$) as the Lagrange-multiplier field enforcing the density constraint, along with the Lagrange multipliers ($\epsilon_i$) that enforce the normalization condition on the GKS orbitals. In other words, we optimize the following Lagrangian with respect to $\Phi$ and $\epsilon_i$,

\begin{equation}
\begin{split}
\mathcal{L} =& 
\Big\langle \Phi \Big| - \sum_i \frac{1}{2} \nabla_i ^2  + \alpha  \sum_{i < j} \frac{1}{|\br_i - \br_j |} \Big| \Phi \Big\rangle + \int  \vloca(\br) \left(\rho(\br) - \rhod(\br)\right) \dr + \sum_i \epsilon_i\left(\int\psi_i^2(\br)\dr - 1\right)
\end{split}
\label{eq:gksInv_lag}
\end{equation}

with electron density
$\rho(\mathbf{r}) = 2\sum^{N_e/2}_i |\psi_i(\mathbf{r})|^2$.
Optimizing $\mathcal{L}$ with respect to $\psi_i$ yields the GKS eigenvalue problem
\begin{equation}
\HGKS \psi_i (\br) = \epsilon_i \psi_i (\br)\,, \quad \hat{H}_{\text{GKS}} = -\frac{1}{2} \nabla^2 +  \alpha \VX[\{\psi_i\}]  + \alpha \vH[\rho](\br) + \vloca(\br)\,.
\label{eq:gksInv_eigen}
\end{equation}
In the above, $\VX[\{\psi_i\}]$ is the nonlocal exchange operator whose action on a function (say $f(\br)$) is given as
\begin{equation} \label{eq:gksInv_VXAction}
\VX f(\br) = - \sum_j \left(\int \frac{\psi_j(\br')f(\br')}{|\br-\br'|}\dr'\right)\psi_j(\br)\,.
\end{equation}
Upon solving Eq.~\ref{eq:gksInv_eigen}, the derivative of $\mathcal{L}$ with respect to $\vloca$ is given by
\begin{equation}
\frac{\delta \mathcal{L}}{\delta \vloca(\br)} = \rho(\br) - \rhod(\br)\,.
\label{eq:gksInv_updatevc}
\end{equation}

Thus, the density mismatch provides the gradient for updating $\vloca$. The update for $\vloca$ in Eq.~\ref{eq:gksInv_updatevc} is structurally analogous to the Wu–Yang inversion strategy  ~\cite{wu2003direct} in KS theory, generalized here to a GKS auxiliary system through the explicit presence of the $\alpha \VX$ operator and the corresponding $\alpha$-dependent local multiplicative potential. Upon convergence, we express $\vloca(\br) = \vext(\br) + (1-\alpha)\vH[\rhod(\br) ] + \vxcalpha(\br)$, where $\vext(\br)=-\sum_I \frac{Z_I}{|\br-\bR_i|}$ is the nuclear potential and $\vH[\rhod](\br)=\int \frac{\rhod(\br')}{|\br-\br'|}\dr'$ is the Hartree potential of $\rhod$. The above form of $\vloca$ results in $\HGKS=-\frac{1}{2}\nabla^2 + \alpha \VX\left[\{\psi_i\}\right]\vext + \vH[\rhod] + \vxcalpha$, and hence, allows us to extract the exact $\alpha$-dependent XC potential for $\rhod$.

\noindent We note that the GKS eigenvalue problem (Eq.~\ref{eq:gksInv_eigen}) is a nonlinear problem, as the input to the GKS Hamiltonian ($\HGKS$) depends on the solution $\psi_i$'s. 
Thus, the GKS eigenvalue problem needs to be solved self-consistently with respect to the $\psi_i$'s. We employ a nested optimization strategy. We start with a trial $\vxcalpha$, and self-consistently solve Eq.~\ref{eq:gksInv_eigen} to  obtain $\psi_i$ and $\rho(\br)$, and then update $\vxcalpha$ using Eq.~\eqref{eq:gksInv_updatevc}. This process is repeated until $\rho(\br)$ matches $\rhod(\br)$ to a prescribed tolerance. We refer the readers to the bottom panel of Fig. \ref{fig:gks_schema} for an overview of the algorithm.

\noindent The resulting $\vxcalpha$ and the corresponding $\psi_i$'s provide reference data to understand and develop  better hybrid XC functionals. By obtaining an accurate reference energy ($\EGS$) for the system from the many-body calculation, the inverse GKS can provide the exact XC energy, as: $ \Exc^\alpha = \EGS - (\Ts[\{\psi_i\}] + \EH[\rhod] + \int \vext(\br) \ \rhod(\br) \ d \br + \EZZ )$, where $\Ts[\{\psi_i\}]=-\sum_{i=1}^{N_e/2}\int \psi_i(\br) \nabla^2 \psi(\br)\dr$ is the GKS non-interacting kinetic energy; $\EH[\rho]=\frac{1}{2}\int\int \frac{\rho(\br)\rho(\br')}{|\br-\br'|}\dr\dr'$ is the Hartree energy; and $\EZZ = \sum_{I=1}^{N_a}\sum_{J=1, J\neq I}^{N_a}\frac{Z_I Z_J}{|\bR_I-\bR_J|}$ is the nuclear-nuclear repulsive energy. Second, through Eq.~\ref{eq:gksInv_Exc}, it also provides the exact remainder XC energy ($\Excalpha$). 

\noindent In the above inverse GKS approach,  the most expensive part is the solution of the nonlinear GKS eigenvalue problem in Eq.~\ref{eq:gksInv_eigen}, for a given $\vxcalpha$. Within the GKS eigenvalue problem, it is action of the exchange operator ($\VX$) on the orbitals ($\psi_i$) that is the most dominant cost, owing to its nonlocal nature. To alleviate this high cost, we employ an adaptively compressed exchange (ACE) approximation to the action of $\VX$. We refer to the Methods section for the details of the ACE approximation, and the overall approach to solve the nonlinear GKS eigenvalue problem, which together accelerate the inverse GKS calculations.

\subsection*{Verification with LDA correlation}

\begin{figure}[htbp!]
    \centering
    \includegraphics[scale=0.4]{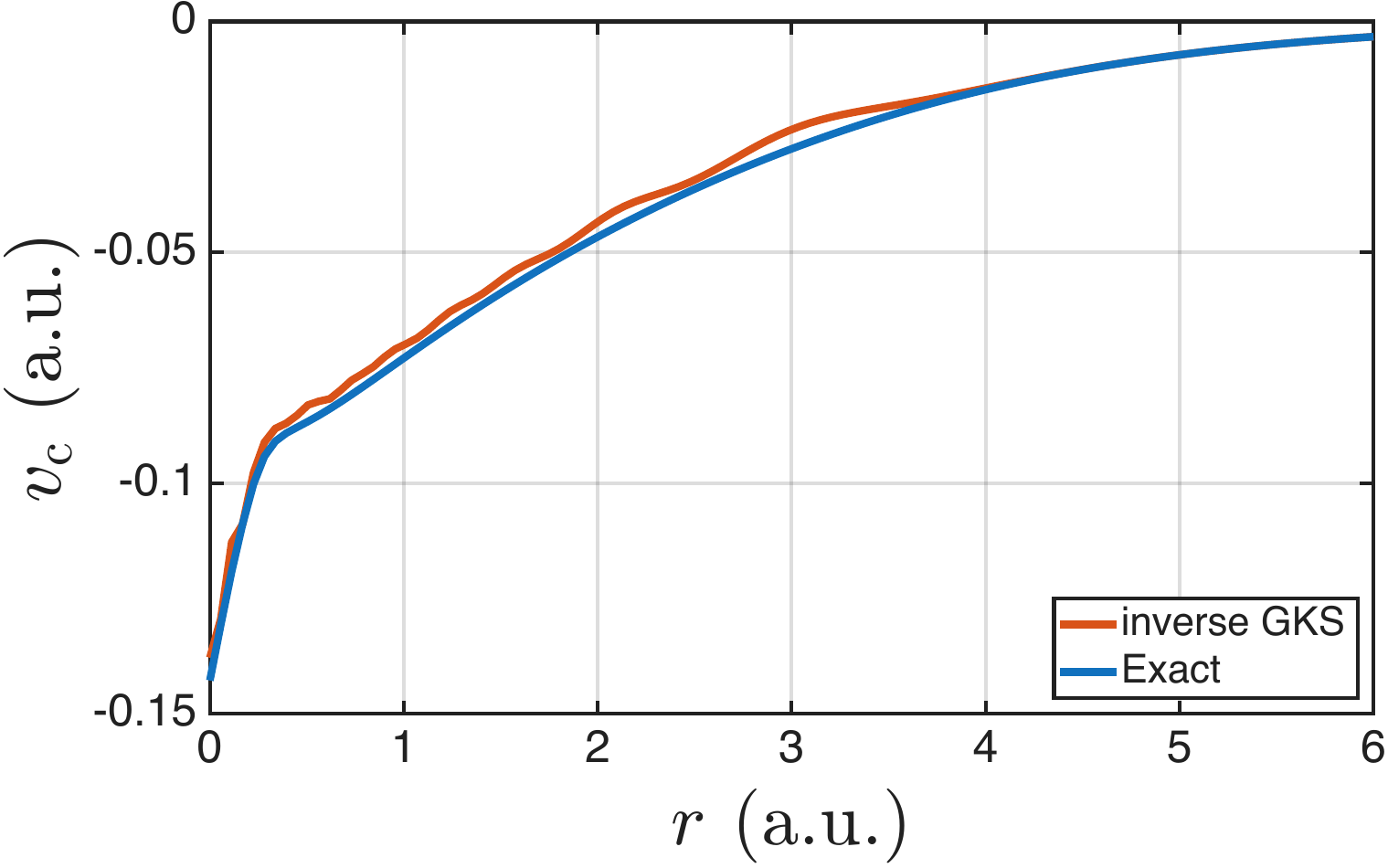}
    \caption{LDA validation test: Comparison of inverse GKS reconstructed and exact potential for a groundstate density obtained using exact exchange + LDA (PW92) correlation.}
    \label{fig:Ne_lda_c_inversion} 
\end{figure}

To assess the accuracy of the algorithm, we consider $\rhod$ for the neon (Ne) atom obtained from a groundstate DFT calculation with full exact exchange (i.e., $\alpha=1$) and the PW92 LDA correlation functional~\cite{PhysRevB.45.13244}, solved using an FE basis. This test allows us to directly compare the $\vxcalpha$ obtained from the inverse GKS calculation with the PW92 correlation potential for $\rhod$ (i.e. $v_{\mathrm{c}}^{\mathrm{LDA}}[\rhod]$). Although seemingly trivial, numerically it represents a difficult test, given the ill-posedness of the discrete inverse DFT problem. To elaborate, past efforts at similar LDA tests for the relatively easier inverse KS problem had suffered from non-unique solutions and/or spurious oscillations~\cite{Gaiduk2013removal}, owing to the finite nature of the atomic orbital (AO) basis employed. The inverse GKS calculation, solved as a 3D problem, is performed until the $L_2$ norm of the error in density ($||\rho-\rhod||_{L_2}=\sqrt{\int\left(\rho(\br)-\rhod(\br)\right)^2\dr}$) is driven below $10^{-5}$.  Figure~\ref{fig:Ne_lda_c_inversion} compares $v_{\mathrm{c}}^{\mathrm{LDA}}[\rhod]$ with the $\vxcalpha$ obtained from inverse GKS calculation on $\rhod$. As evident, the solution from inverse GKS is in good agreement with $v_{\mathrm{c}}^{\mathrm{LDA}}[\rhod]$ and exhibits no spurious oscillations at the scale of the potential features underscoring the reliability of the proposed inverse GKS method.

\subsection*{Exact GKS exchange-correlation potentials}
We evaluate the exact exchange-correlation potentials corresponding to accurate groundstate densities obtained from configuration interaction (CI) calculations. These calculations in general, use atomic orbital (AO) basis, such as  Gaussian- or Slater-type orbitals.  However, as noted in earlier works~\cite{kanungo2019exact, Kanungo2023, Kanungo2025accelerating}, AO basis sets, owing to their finite (incomplete) nature, do not accurately capture the asymptotics of the density near the nuclei and in the far-field (low density region). Gaussian basis sets, in particular, lack the nuclear cusp and exhibit Gaussian rather than exponential decay in the far-field. These inaccurate asymptotics can induce unphysical oscillations in the $\vxcalpha$ obtained through inverse GKS, similar to those observed in inverse KS~\cite{kanungo2019exact, Kanungo2025accelerating}. Slater-type basis sets, by contrast, are better suited to capture these asymptotics, and have been shown to alleviate such issues in the context of Ryabinkin-Kohut-Staroverov (RKS) method for inverse KS~\cite{tribedi2023exchange}. However, minor oscillations may persist depending on the choice of basis, especially when used with a complete basis like finite-elements for the inverse problem~\cite{Kanungo2025accelerating}.
 To alleviate these artifacts, we employ the following two strategies. First, we add a small correction, $\Delta \rho(\br)$, to the $\rhod(\br)$, given as
\begin{equation} \label{eq:gksInv_deltarho}
    \Delta \rho(\br) = \rho_{\text{DFA}}^{\text{FE}}(\br) - \rho_{\text{DFA}}^{\text{AO}} (\br)\,. 
\end{equation}
In the above, $\rho_{\text{DFA}}^{\text{FE}}$  refers to the self-consistent density obtained using a systematically convergent FE basis with an approximate density functional. $\rho_{\text{DFA}}^{\text{AO}}(\br)$ refers to the self-consistent density using the same approximate density functional, albeit solved using the AO basis used in the CI calculation. Broadly, $\Delta \rho$ captures the AO basis set error near the nuclei, and hence, helps in alleviating any basis set artifact in $\vxcalpha$ near the nuclei. We refer to Refs~\cite{kanungo2019exact, Kanungo2025accelerating} for the efficacy of the $\Delta \rho$ correction, in the context of inverse KS. We have shown that the accuracy of the inverse calculations is insensitive to the choice of functional used to compute the $\Delta\rho$ correction in our previous works~\cite{kanungo2019exact, subramanian2026invdft}.  Second, to alleviate any spurious behavior in $\vxcalpha$ stemming from the incorrect decay in $\rhod$, we enforce the known $-(1-\alpha)/r$ decay on the $\vxcalpha$ in the low density regime. Specifically, we set $\vxcalpha = -\frac{(1-\alpha)}{N_e}\vH[\rhod]$ in regions where ($\rhod < 10^{-6}$). 

\begin{figure}[htbp!]
    \centering
    \includegraphics[scale=0.4]{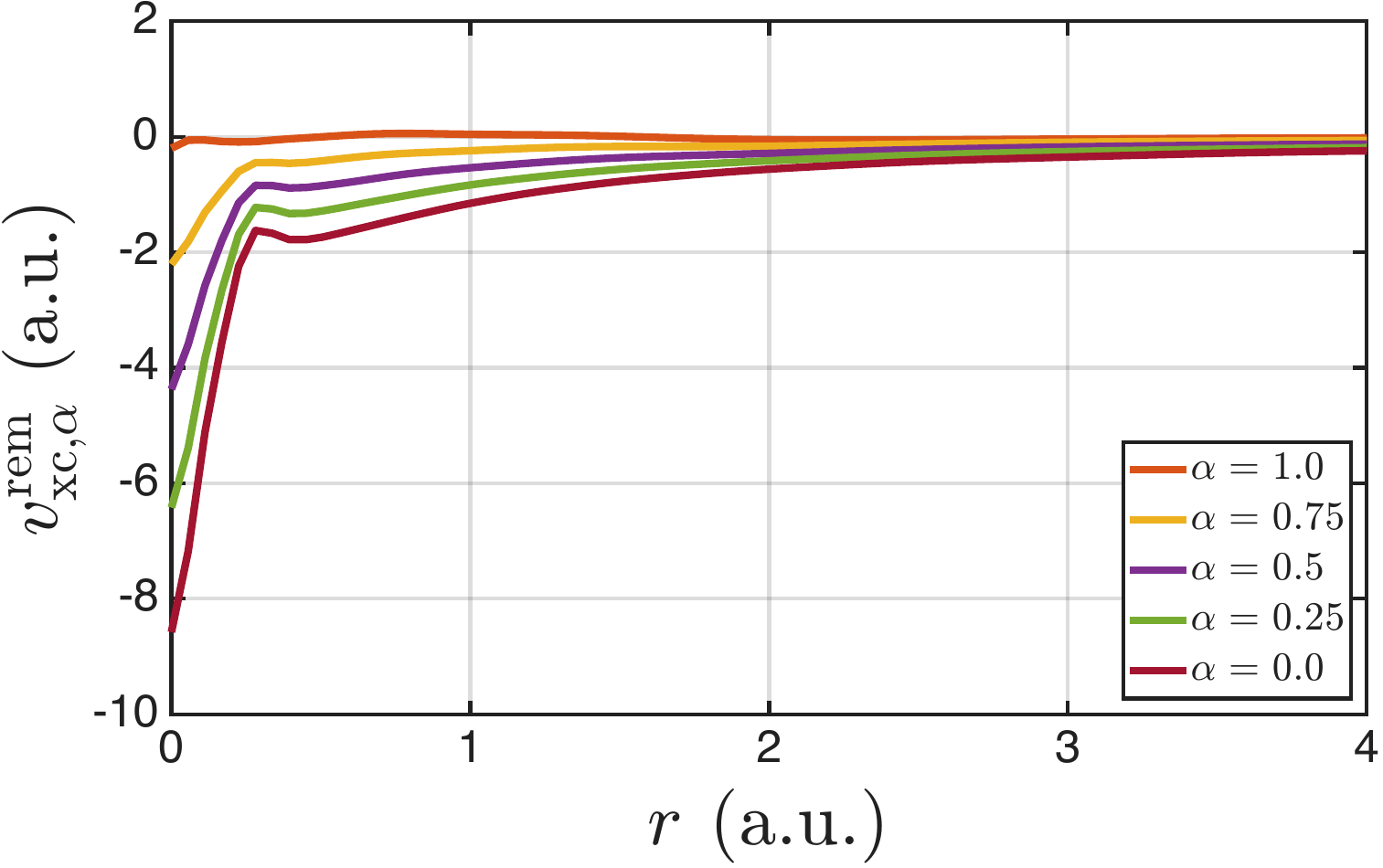}
    \caption{$\vxcalpha$ potentials for different fractions ($\alpha$) of exact exchange for a Ne atom.}
    \label{fig:ne_gks_inversion}
\end{figure}

\begin{figure}[htbp!]
    \centering
    \includegraphics[scale=0.4]{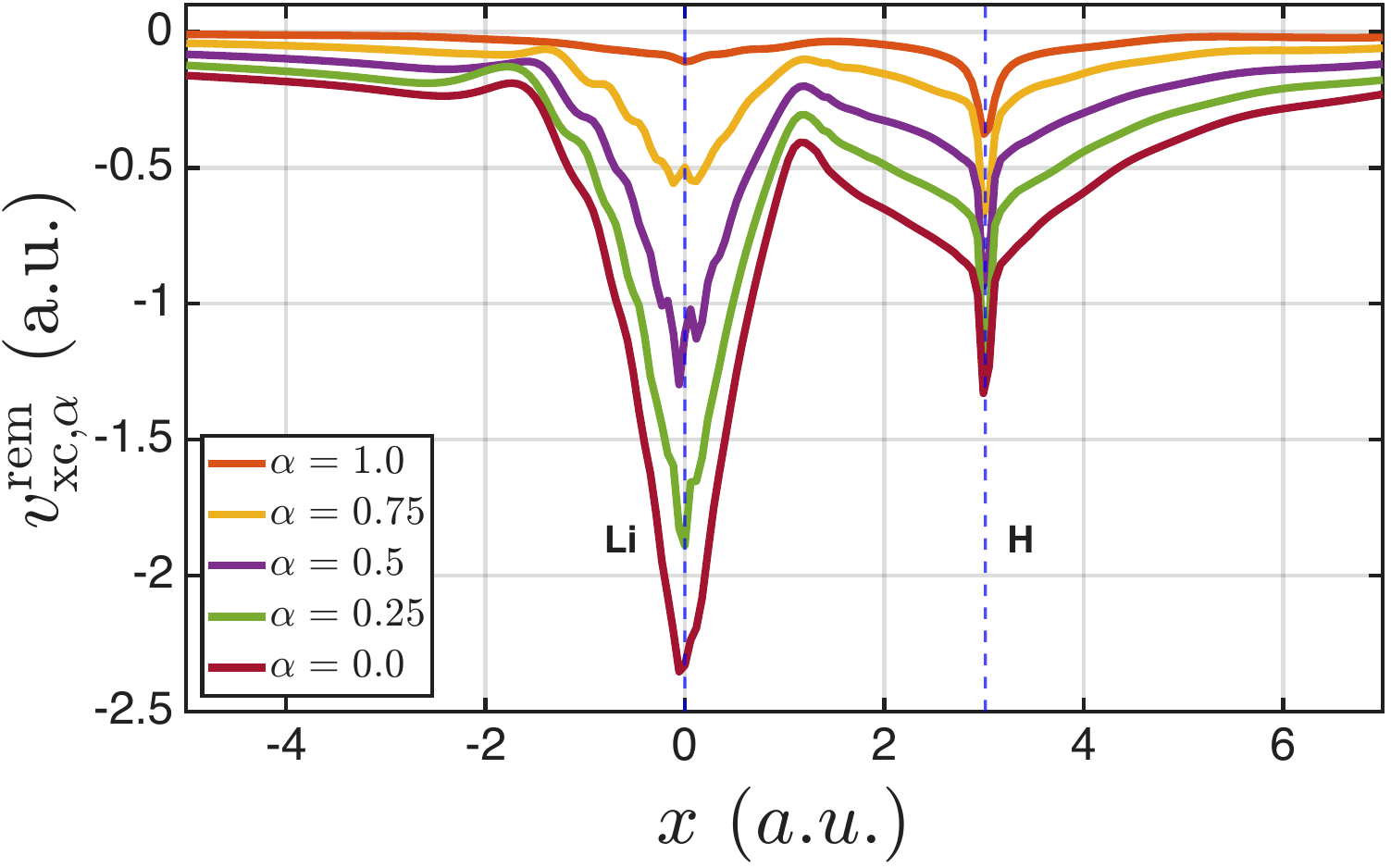}
    \caption{$\vxcalpha$ potentials for different fractions ($\alpha$) of exact exchange for LiH molecule.}
    \label{fig:lih_gks_inversion}
\end{figure}

\begin{figure}[htbp!]
    \centering
    \includegraphics[scale=0.4]{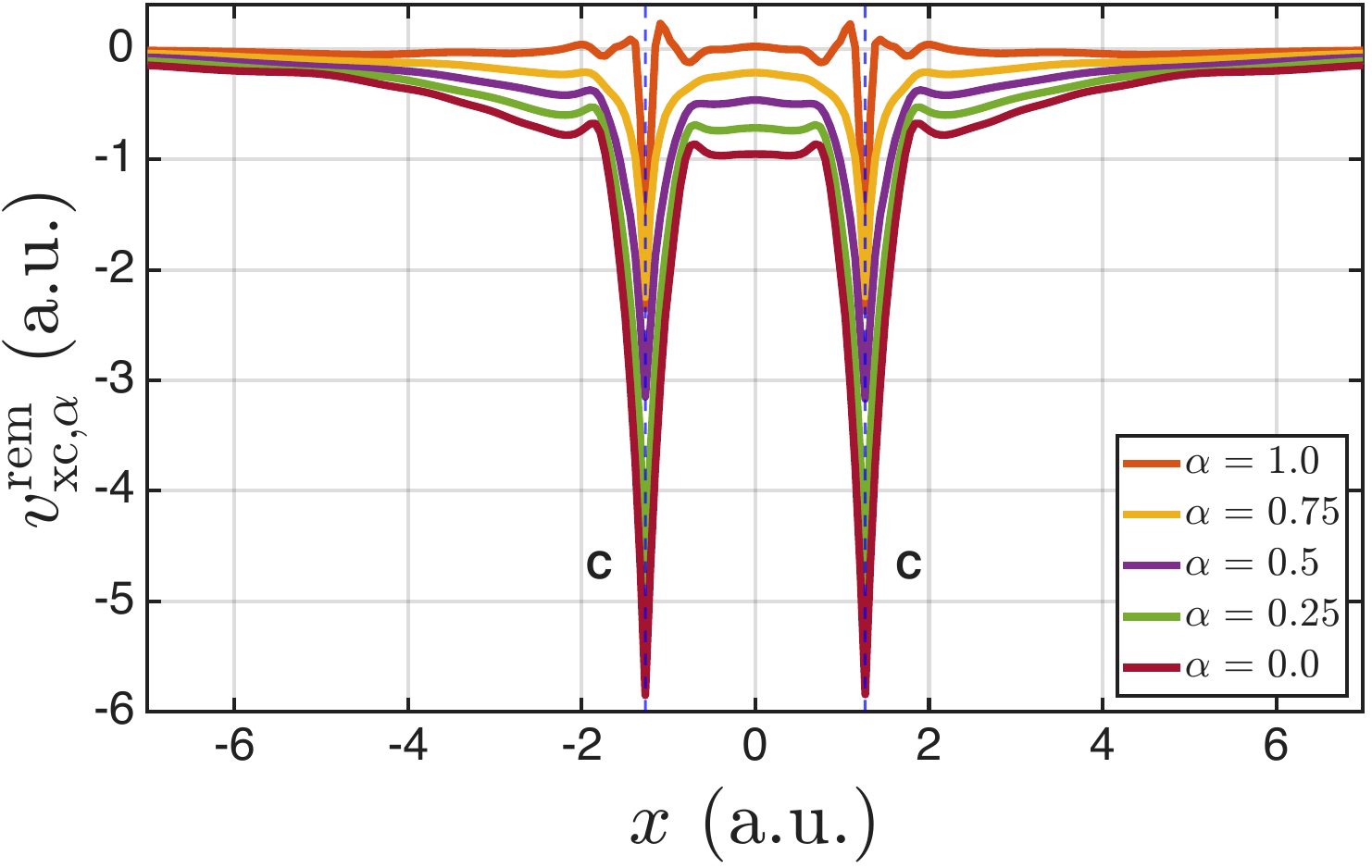}
    \caption{ $\vxcalpha$ potentials for different fractions ($\alpha$) of exact exchange for $ \mathrm{C}_2 \mathrm{H}_4$ molecule.}
    \label{fig:c2h4_gks_inversion}
\end{figure}

\begin{figure}[htbp!]
    \centering
    \includegraphics[scale=0.7]{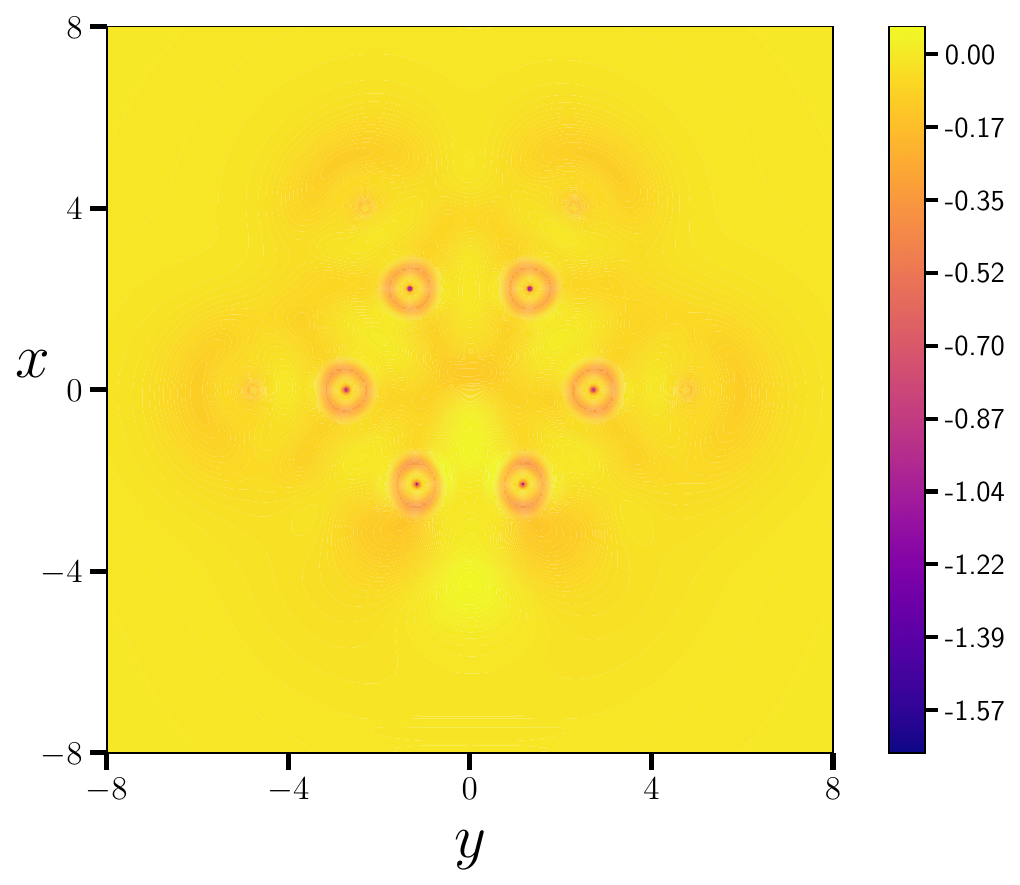}
    \caption{$v_{\mathrm{xc}, \alpha=1}^{\mathrm{rem}}$  for the benzyne molecule ($\text{C}_6\text{H}_4$), shown on the plane of the molecule.}
    \label{fig:benzyne_gks_alpha1p0_inversion}
\end{figure}

We present the exact $\vxcalpha$ and energies for different benchmark systems, each at five different values of $\alpha=\{0, 0.25, 0.5, 0.75, 1.0\}$. The energies we consider include the non-interacting kinetic energy ($\Ts$), the electrostatic energy ($\Ees[\rho] = \EH[\rho] + \int \rho \vext \dr + \EZZ$, where $\EZZ$ is the nuclear-nuclear repulsive energy), and the exact exchange energy ($\Exx$) for these systems.  The $\vxcalpha$ for $\alpha=0$ represents the Kohn-Sham XC potential and that for $\alpha=1$ represents the exact GKS correlation potential. All inverse GKS calculations are deemed converged when  $||\rho-\rhod||_{L_2} < 10^{-3}$.  Figure~\ref{fig:ne_gks_inversion} shows the $\vxcalpha$ for different $\alpha$ for the Ne atom. As evident, $\vxcalpha$ decreases by an order of magnitude with increasing $\alpha$. The intershell structure in $\vxcalpha$ (the hump in the potential around $r=0.3$) becomes less prominent with increasing $\alpha$. This indicates the intershell features to be largely a consequence of the exchange operator. We next conduct a similar analysis for molecules. Figure~\ref{fig:lih_gks_inversion} and Fig.~\ref{fig:c2h4_gks_inversion} present the $\vxcalpha$ for the LiH and $\text{C}_2\text{H}_4$ molecules, respectively, each at their equilibrium geometry. Similar to Ne atom, the magnitude of $\vxcalpha$ and the intershell structure diminishes with increasing $\alpha$. Figure~\ref{fig:benzyne_gks_alpha1p0_inversion} shows the $\vxcalpha$ for benzyne ($\text{C}_6\text{H}_4$), which is a strongly correlated system. Given the high computational cost associated with inverse GKS for larger molecules, we limit the evaluation of $\vxcalpha$ for benzyne to only $\alpha=1$. As evident, the $v_{\mathrm{xc}, \alpha=1}^{\mathrm{rem}}$ for benzyne is devoid any numerical artifacts. These results, spanning both weakly and strongly correlated systems, highlight the accuracy, robustness, and scalability of our inverse GKS method. We also report the energies ($\Ts + \Ees$ and $\Exx$ ) for these systems in Table \ref{tab:gks_energy_all}. We find that, at least for these systems, the energies are largely insensitive to the choice of $\alpha$, consistent with the conventional wisdom that the KS and GKS density matrices are nearly identical~\cite{Cuevas2015}, at least by energy metrics. However, in the next section we provide examples for which this conventional wisdom does not hold. The molecular geometries, the CI density matrices, and the corresponding AO basis sets used in these calculations are provided in the SI.

\begin{table}[htbp!]
\centering
    \caption{Energies for GKS potentials for Ne, LiH(eq), $\text{C}_2\text{H}_4$ and benzyne (C$_6$H$_4$). $\Ts$, $\Ees$, and $\Exx$ denote the non-interacting kinetic energy, the electrostatic energy, and the exact exchange energy, respectively. The energies are in Ha.}
    \label{tab:gks_energy_all}
    \begin{tabular}{|c|c|c|c|c|c|c|c||}
    \hline
    & & $\alpha = 0.0$ & $\alpha = 0.25$ & $\alpha = 0.5$ & $\alpha = 0.75$ & $\alpha = 1.0$  \\
    \cline{1-7}
    Ne & $\Ts + \Ees$ & -116.4681 & -116.4681 & -116.4679 & -116.4666 & -116.4657 \\
    \cline{2-7}
    & $\Exx$ & -12.0709 & -12.07544 & -12.07616 & -12.07775 & -12.07875 \\
    \hline
     LiH(eq) & $\Ts + \Ees$ & -5.8423 & -5.8421 &	-5.8421	& -5.842 & -5.8418 \\
    \cline{2-7}
    & $\Exx$ &  -2.1432 & -2.1447 & -2.1449 & -2.1450 & -2.1452 \\
    \hline
    $\text{C}_2\text{H}_4$ & $\Ts + \Ees$ & -66.3325 &	-66.3322	& -66.3317	& -66.3284 &	-66.3258 \\
    \cline{2-7}
    & $\Exx$ & -11.7246 & -11.7279 & -11.7303 &  -11.7348 & -11.7379\\
    \hline
    Benzyne & $\Ts + \Ees$ & -- & -- & -- &  -- & -196.9517 \\
    \cline{2-7}
    & $\Exx$ & --  & -- &  -- &  -- & -32.5065 \\
    \hline
    \end{tabular}
\end{table}

\subsection*{Assessing common approximations to $v^{\mathrm{rem}}_{\mathrm{xc}, \alpha}$}

Given the computational and numerical difficulty of inverse GKS, over the years, approximations to $\vxcalpha$ based on quantities available from inverse KS calculations have been pursued. One such approximation, based on perturbative analysis, is \cite{garrick2020exact,trushin2025accurate,Cuevas2015} 
\begin{equation}
    \vxcalpha \approx  (1 - \alpha) v_{\mathrm{x}}^{\mathrm{KS}} +  v_{\mathrm{c}}^{\mathrm{KS}}\,,
    \label{eq:vxc_linear_alpha}
\end{equation}
where $v_{\mathrm{x}}^{\mathrm{KS}}$ and $v_{\mathrm{c}}^{\mathrm{KS}}$ denote the Kohn-Sham exchange and correlation potentials, respectively. The $v_{\mathrm{x}}^{\mathrm{KS}}=\frac{\partial E_{\mathrm{xx}}^{\mathrm{KS}}}{\partial \rho(\br)}\Big|_{\rhod}$, where $E_{\mathrm{xx}}^{\mathrm{KS}}$ denotes the exchange energy evaluated using the exact Kohn-Sham orbitals (i.e., orbitals corresponding to $v_{\mathrm{xc},\alpha=0}^{\mathrm{rem}}$). Accordingly, $v_{\mathrm{c}}^{\mathrm{KS}} = \vxc - v_{\mathrm{x}}^{\mathrm{KS}}$, where $\vxc = v_{\mathrm{xc},\alpha=0}^{\mathrm{rem}}$ is the exact Kohn-Sham XC potential. Evaluation of $v_{\mathrm{x}}^{\mathrm{KS}}$ requires solving the optimized effective potential (OEP) integral equations in a complete basis ~\cite{Sharp1953variational, Talman1976}, which presents a much more challenging problem than inverse KS. However, it has been shown, both theoretically and numerically~\cite{Filippi1996separation, Ryabinkin2013accurate, Kohut2014hierarchy, Cuevas2015, trushin2025accurate}, that the XC potential corresponding to the self-consistent Hartree-Fock density can serve as a good approximation to $v_{\mathrm{x}}^{\mathrm{KS}}$, i.e., $\vxc[\rho_{\mathrm{HF}}]\approx v_{\mathrm{x}}^{\mathrm{KS}}[\rhod]$. In this work, we use this Hartree-Fock based approximation to  $v_{\mathrm{x}}^{\mathrm{KS}}$ and evaluate 
\begin{equation} \label{eq:vcksapprox}
v_{\mathrm{c}}^{\mathrm{KS}}[\rhod] \approx \vxc[\rhod] - \vxc[\rho_{\mathrm{HF}}]\,.
\end{equation}

\begin{figure}[htbp!]
    \centering
    \includegraphics[scale=0.4]{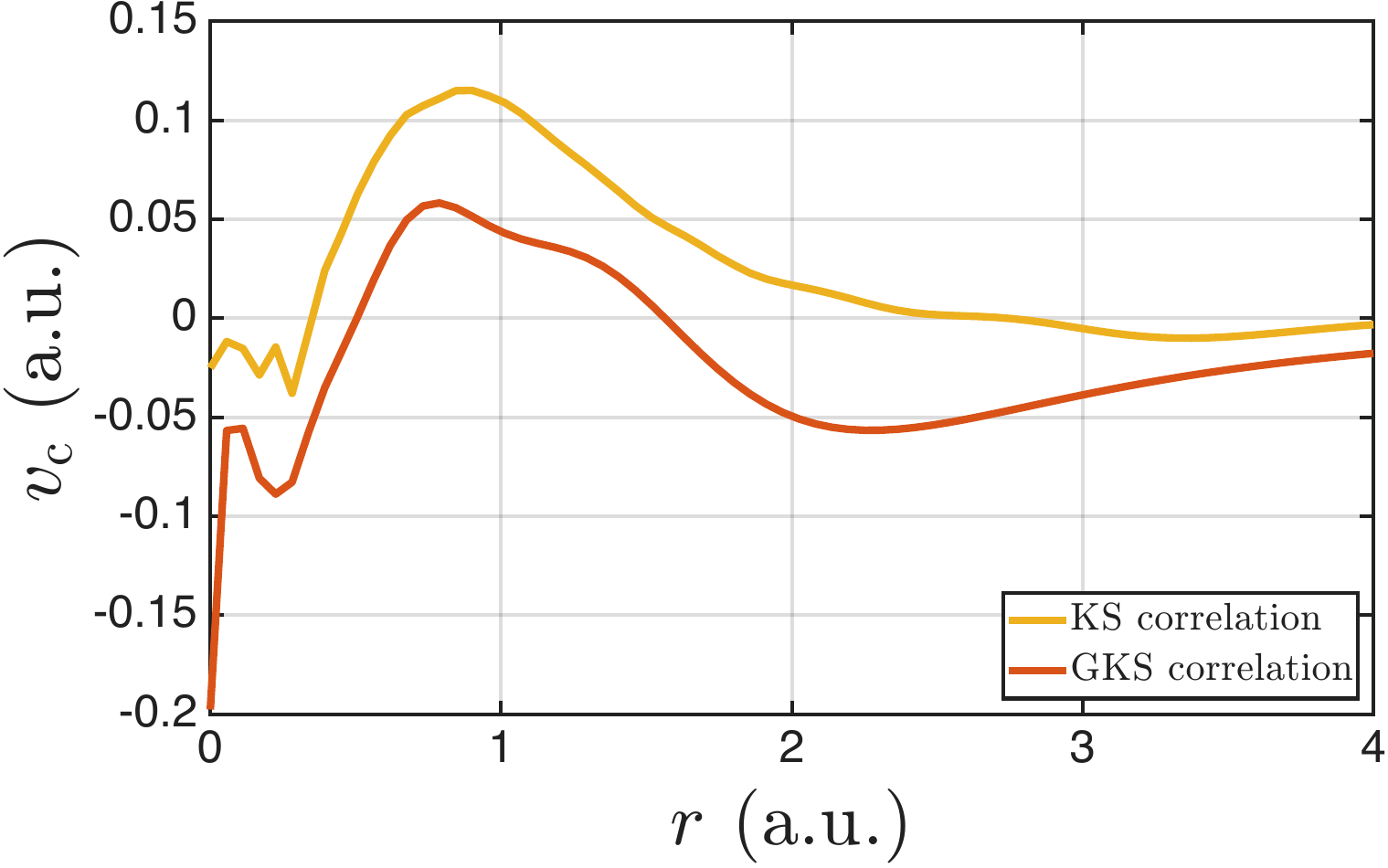}
    \caption{Comparison of GKS and KS correlation potentials for neon (Ne) atom.}
    \label{fig:Ne_gks_ks_vc}
\end{figure}

\begin{figure}[htbp!]
\centering
\begin{subfigure}{0.5\textwidth}
    \centering
    \includegraphics[scale=0.3]{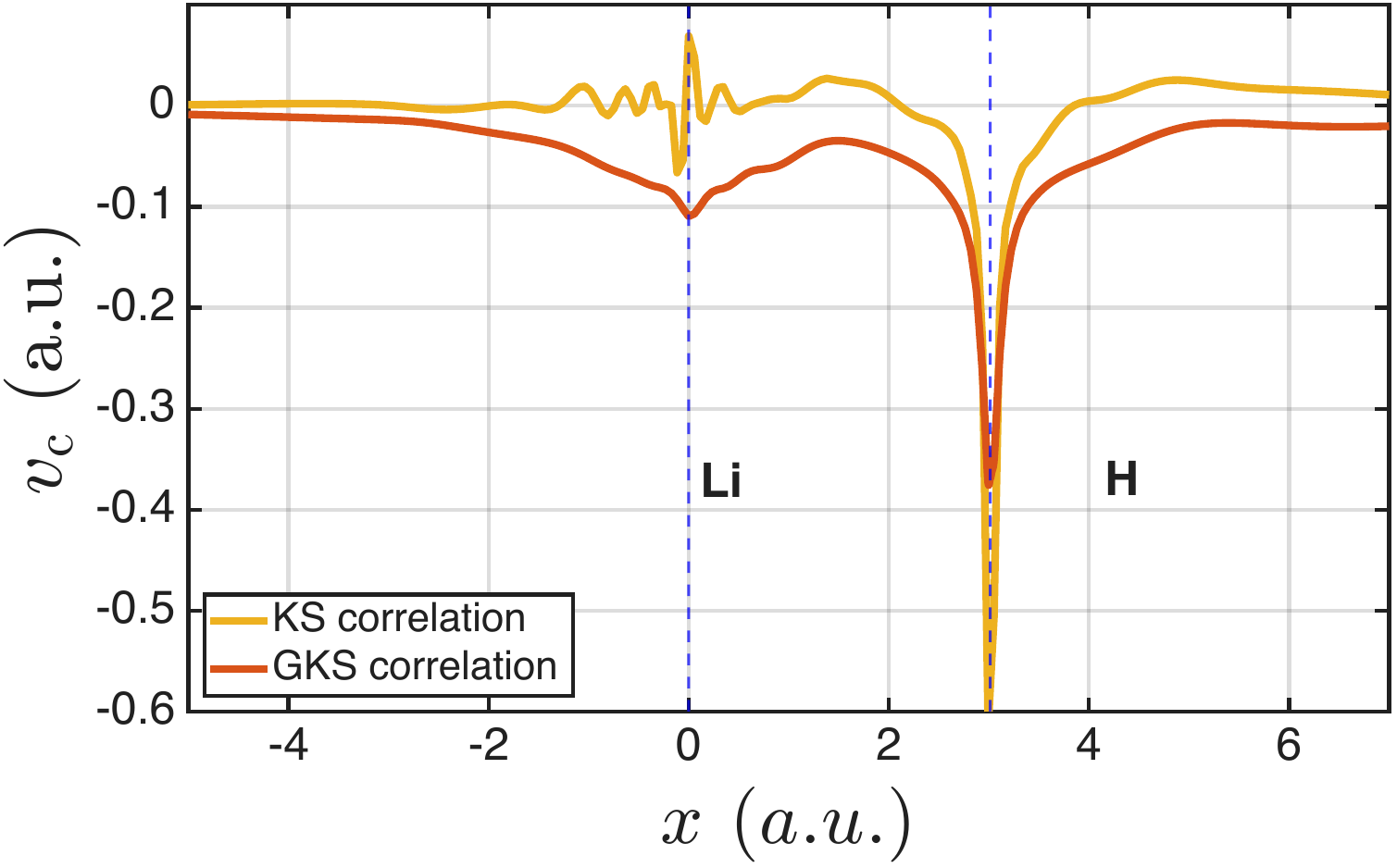}
\end{subfigure}%
\begin{subfigure}{0.5\textwidth}
    \centering
     \includegraphics[scale=0.3]{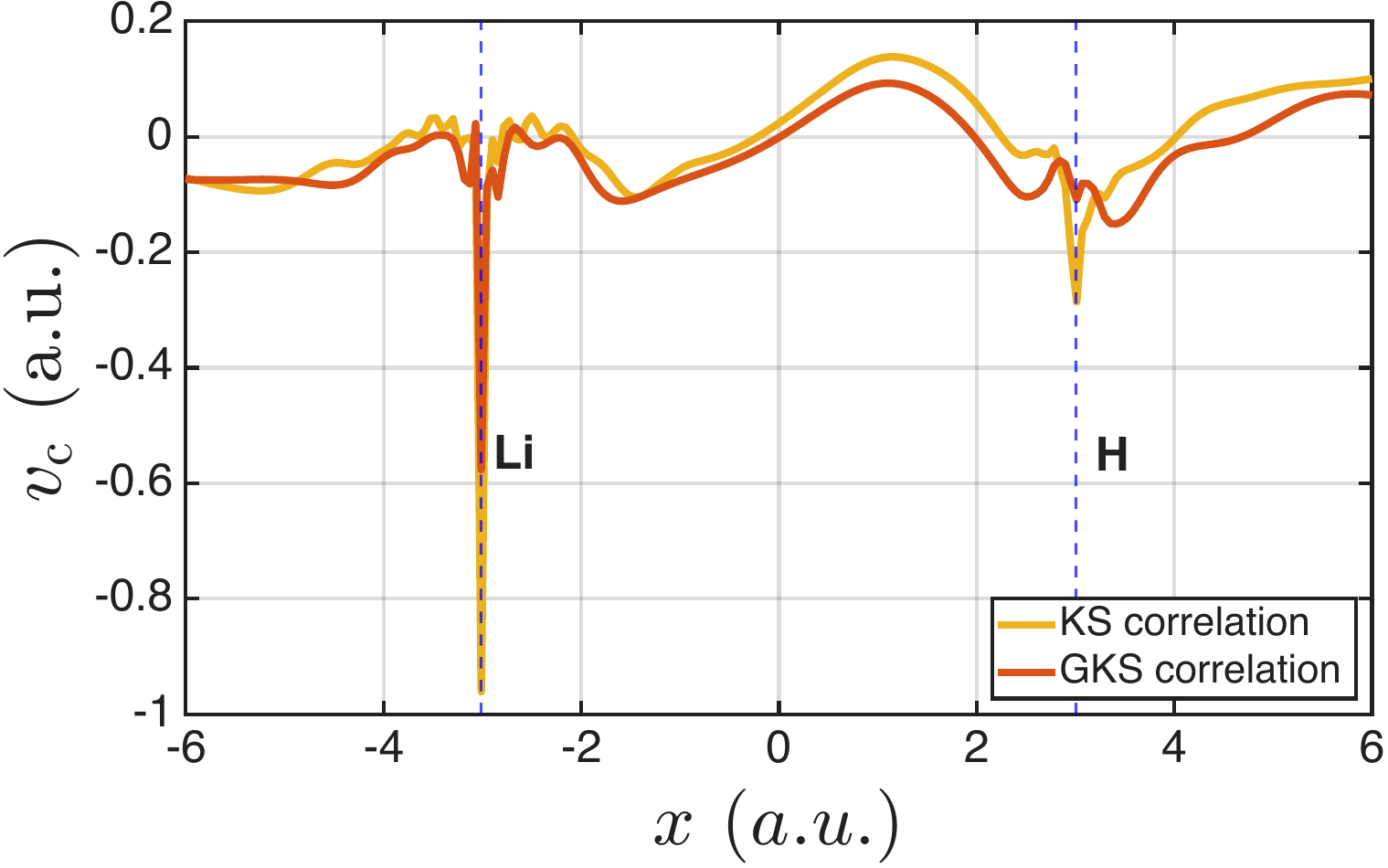}
\end{subfigure}
\caption{Comparison of KS and GKS correlation potentials for LiH molecules, along the bond. Left: LiH(eq). Right: LiH(2eq).}
\label{fig:LiH_eq_2eq}
\end{figure}

\begin{figure}[htbp!]
\centering
\begin{subfigure}{0.5\textwidth}
\centering
    \includegraphics[scale=0.3]{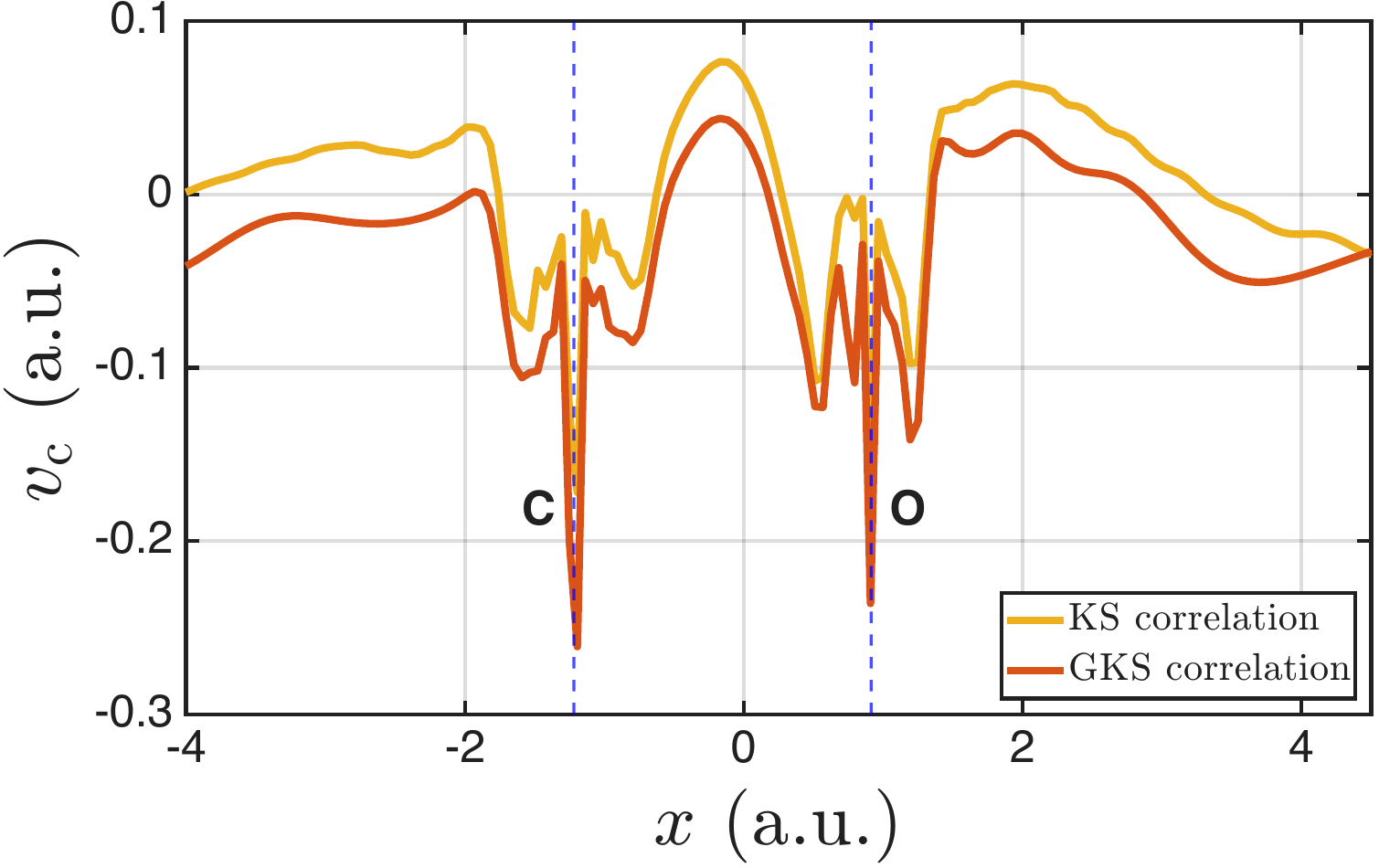}
\end{subfigure}%
\begin{subfigure}{0.5\textwidth}
\centering
    \includegraphics[scale=0.3]{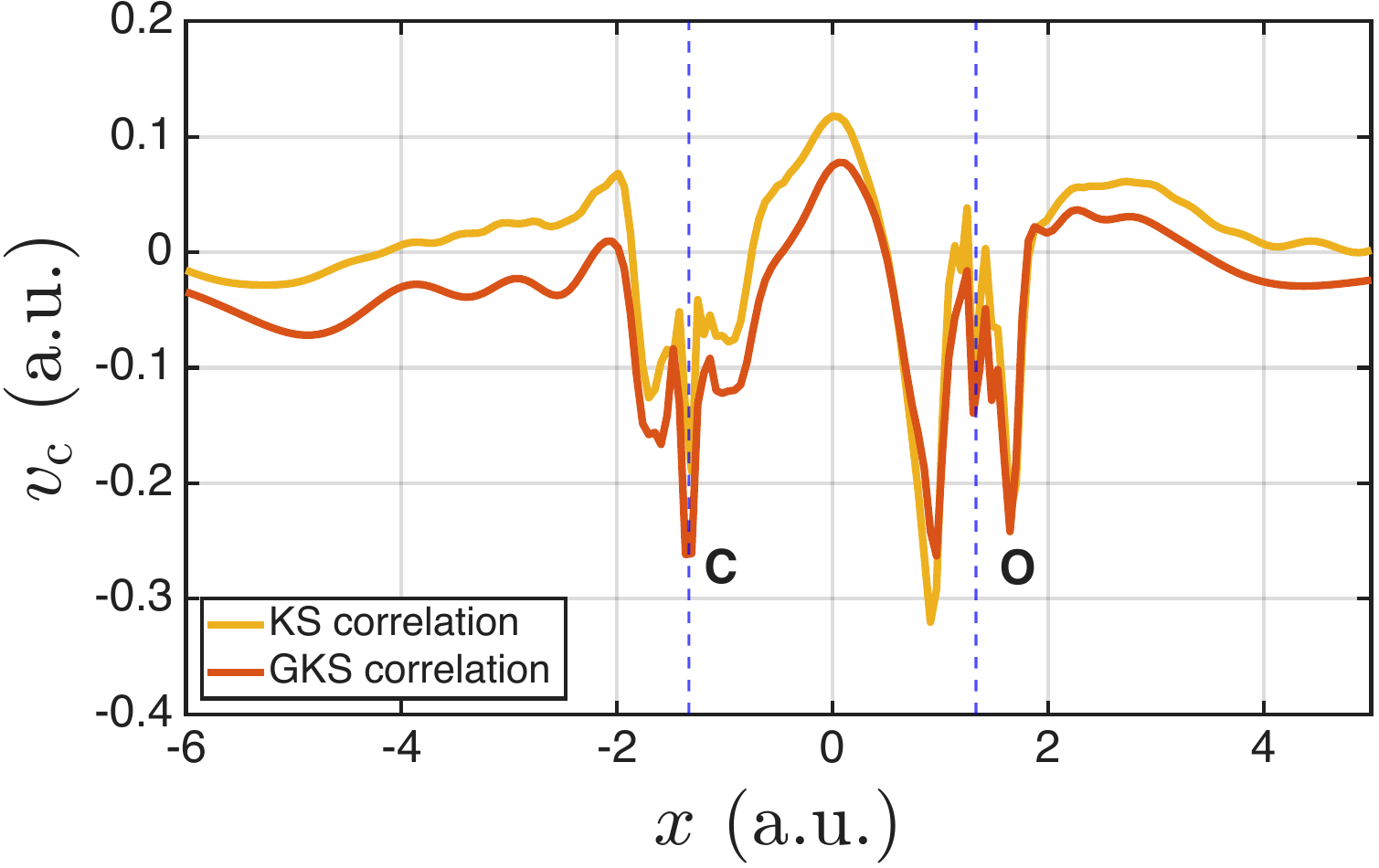}
\end{subfigure}%
\caption{Comparison of KS and GKS correlation potentials for CO molecules, along the bond. Left: CO(eq). Right: CO(1.25eq).} 
\label{fig:CO_eq_1.25eq}
\end{figure}

Using our inverse GKS method, we test the approximation in Eq.~\ref{eq:vxc_linear_alpha} for $\alpha=1$. That is, we check if the exact GKS correlation potential ($v_{\mathrm{c}}^{\mathrm{GKS}} = v_{\mathrm{xc}, \alpha=1}^{\mathrm{rem}}$) and KS correlation potential ($v_{\mathrm{c}}^{\mathrm{KS}}$) are close to each other. Figure~\ref{fig:Ne_gks_ks_vc} compares the $v_{\mathrm{c}}^{\mathrm{GKS}}$ and $v_{\mathrm{c}}^{\mathrm{KS}}$ for Ne atom. Figure~\ref{fig:LiH_eq_2eq} offers the same comparison for LiH  molecule at both equilibrium and twice the equilibrium bond-length, hereafter denoted as LiH(eq) and LiH(2eq). Similarly, Fig.~\ref{fig:CO_eq_1.25eq} presents the comparison for CO molecule at equilibrium and 1.25$\times$ the equilibrium bond-length, hereafter denoted as CO(eq) and CO(1.25eq). For all the systems, the GKS and KS correlation potential differ both qualitatively and quantitatively. Beyond the visual comparison, a more instructive measure to compare the two potentials is through various single-electron energies. Specifically, we compare the $\Ts$, $\Ees[\rho]$, $\Exx$, and the mean-field energy ($\EMF = \Ts + \Ees + \Exx$), evaluated using the orbitals ($\psi_i$) corresponding to $v_{\mathrm{c}}^{\mathrm{GKS}}$ and $v_{\mathrm{c}}^{\mathrm{KS}}$. We obtain the orbitals by self-consistently solving the GKS eigenvalue problem in Eq.~\ref{eq:gksInv_eigen} with $v_{\mathrm{xc},\alpha=1}^{\mathrm{rem}}$ set to $v_{\mathrm{c}}^{\mathrm{GKS}}$ and $v_{\mathrm{c}}^{\mathrm{KS}}$. This measures the differences in energies one would obtain, if the $v_{\mathrm{c}}^{\mathrm{GKS}}$ (which is exact) is replaced by $v_{\mathrm{c}}^{\mathrm{KS}}$ and is used in conjuction with the exact exchange. Table~\ref{tab:gks_ks_comparison} compares the energies for Ne, LiH(eq), LiH(2eq), CO(eq), and CO(1.25eq). The stretched molecules---LiH(2eq) and CO(1.25eq)---present strongly correlated systems. We observe that for Ne, LiH(eq), and CO(eq), the $\EMF$ from GKS and KS correlation potentials are in good agreement ($< 2$ mHa difference). Indeed, for LiH(eq) and CO(eq), even the individual energies ($\Ts$, $\Ees$, and $\Exx$) are in good agreement. However, this changes for the stretched molecules (LiH(2eq) and CO(1.25eq)), where the $\Exx$ energies show notable differences ($>20$ mHa). Overall, the $\EMF$ for the stretched molecules differ by more than 11 mHa. Thus, from an energy perspective, we observe that, while the GKS and KS correlation potential are agree with each other for weakly correlated systems (i.e., Ne, LiH(eq), and CO(eq)), they differ significantly (in relation to the desired chemical accuracy of 1 mHa) for strongly correlated systems (e.g. LiH(2eq) and CO(1.25eq)). Based on these results, we hypothesize Eq.~\ref{eq:vxc_linear_alpha} to be valid for weakly correlated systems, but not necessarily for strongly correlated ones.  

\begin{table}[htbp!]
    \caption{Comparison of energies for GKS and KS correlation potentials when used with full  exact exchange across different systems. $\Ts$, $\Ees$, and $\Exx$ denote the non-interacting kinetic energy, the electrostatic energy, and the exact exchange energy, respectively. $\EMF = \Ts +\Ees + \Exx$.}
    \label{tab:gks_ks_comparison}
    \begin{tabular}{|c||c|c|c|c|c|c||c|}
    \hline
    System & \multicolumn{2}{c|}{$\Ts$ (Ha)} & \multicolumn{2}{c|}{$\Ees$ (Ha)} & \multicolumn{2}{c||}{$\Exx$ (Ha)} & Diff. $\EMF$ (mHa) \\
    \hline
    & GKS & KS & GKS & KS & GKS & KS & (GKS - KS) \\
    \hline
    Ne & 128.6005 & 128.5817 & -245.0662 & -245.0514 & -12.0788 & -12.0746 & -0.15 \\
    \hline
    LiH(eq) & 7.999 & 8.000 & -13.841 & -13.842 & -2.145 & -2.144 & 0.073 \\
    \hline
    LiH(2eq) & 7.799 & 7.787 & -13.691 & -13.692 & -2.004 & -1.979 & 11.50 \\
    \hline
    CO(eq) & 112.823 & 112.824 & -212.286 & -212.291 & -13.317 & -13.312 & -2.277 \\
    \hline
    CO(1.25eq) & 112.060 & 112.059 & -211.651 & -211.658 & -13.043 & -13.021 & -13.30 \\
    \hline\hline
    \end{tabular}
\end{table}

\section{Discussion}

We have presented an accurate method  to evaluate the exact local multiplicative potential ($\vxcalpha$) from a target density, within the generalized-Kohn-Sham (GKS) formulation. The key ingredients in our approach are: (i) the use of systematically convergent finite-element (FE) basis, by which the inverse GKS problem can be solved robustly; (ii) use of asymptotic corrections to the target density and enforcement of appropriate boundary conditions on $\vxcalpha$ to avoid numerical artifacts; and (iii) use of adaptively compressed exchange (ACE) approximation to the exchange operator ($\VX$) to substantially accelerate its action on orbitals. We have demonstrated the efficacy of our approach by obtaining the exact $\vxcalpha$ , for different $\alpha$'s, using accurate configuration interaction (CI) densities, for different molecules, spanning weakly correlated and strongly correlated ones. Through the benzyne molecule, we demonstrated the ability of our proposed method to handle large systems. Overall, to the best of our knowledge, this offers the first attempt at obtaining the $\vxcalpha$'s for molecules. We also demonstrate that a commonly held but previously untested assumption that GKS and KS correlation potentials yield similar energies holds only for weakly correlated systems and breaks down in the strongly correlated regime.  Inverse DFT is increasingly serving as a powerful tool for machine-learning of accurate XC functionals, through the use of exact XC potentials and energies~\cite{Schmidt2019machine, Kanungo2025learning}. However, all such attempts are limited to the KS formalism. This work, by offering exact $\vxcalpha$, exact $\Excalpha$, and the corresponding orbitals ($\psi_i$), paves the way for machine-learning of accurate XC functionals within GKS formalism of DFT which can remedy the notable deficiencies in DFT, of self-interaction and static correlation errors~\cite{Cohen2012}. 

\section*{Methods}
\subsection*{Efficient GKS Eigensolve}\label{sec:efficient}
The most expensive part of the above inverse GKS approach is the solution of the GKS eigenvalue problem in Eq.~\ref{eq:gksInv_eigen} for a given $\vxcalpha$. Being a nonlinear problem, it needs to be solved self-consistently with respect to $\psi_i$'s. We employ a nested self-consistent strategy, where the outer loop updates the exchange operator ($\VX$) and the inner loop updates the orbitals ($\psi_i$). To elaborate, in the outer loop we take a set of input orbitals ($\psiini$) and  define $\VXin=\VX[\{\psiini\}]$. Subsequently, in the inner loop, we solve the GKS eigenvalue problem with the fixed $\VXin$  until self-consistency in the density ($\rho$) is reached. The output orbitals ($\psiouti$) from the inner self-consistent solve serves as the input ($\psiini$) for the next iteration for the outer loop. We repeat the process until self-consistency in $\psi_i$ is achieved, i.e., until the difference in the exact energy ($\Exx$) corresponding to $\psiini$ and $\psiouti$ drops below a prescribed tolerance. We note that the inner self-consistent loop resembles the usual self-consistent field iteration in Kohn-Sham DFT. To elaborate, in the inner loop, we start with an input density $\rhoin$ (defined using $\psiini$) and define its Hartree potential $\vH[\rhoin]$. We then solve the linear GKS eigenvalue in presence of $\VXin$ and $\vH[\rhoin]$ to find the output orbitals $\psiouti$ and its corresponding density $\rhoout$. The $\rhoout$ serves as $\rhoin$ for the next iteration of the inner loop, which is continued until $\rhoin$ and $\rhoout$ match to a chosen tolerance.

\noindent Within the above nested self-consistent approach, the major computational cost stems from the action of $\VXin$ on a set of orbitals $\psi_i$. To elaborate, one needs to solve the inner linear eigenvalue problem (i.e., for a fixed $\VXin$  and fixed $\vH[\rhoin]$) using an iterative eigensolver (e.g., Chebyshev filtering~\cite{zhou2006self} or Jacobi-Davidson~\cite{Sleijpen2000jacobi}). Thus, one has to apply $\VXin$ to a set of orbitals $\psi_i$ at each iteration of the linear eigensolver, which can easily add  to an enormous computational cost. We alleviate this cost by employing the adaptively compressed exchange (ACE) approximation~\cite{lin2016adaptively} to expedite the action of $\VXin$ inside the inner linear eigensolver. ACE is based on the idea that although the exchange-operator ($\VXin$) has full-rank, its action on a set of functions can be approximated using a low-rank operator. To elaborate, given an initial set of functions $\{g_i(\br)\}$ which closely span the space of $\psi_i(\br)$, we define $\VXACE[\{\psiini,g_i\}](\br,\br') = -\sum_k \zeta_k(\br)\zeta_k(\br')$. Here the projection vectors $\zeta_k$ are defined as $\zeta_k(\br)=\sum_l u_l(\br) (\bL^{-1})_{k,l}$, where $u_l(\br)=\VX[\{\psiini\}]g_l(\br)$ and the $\bL$ is the Cholesky factor of the $\bS$ matrix: $S_{i,j}=\int g_i(\br) u_j(\br)\dr$. That is, $\bS = -\bL\bL^T$. The key idea is that as $\psi_i$'s get iteratively updated in the linear eigensolver, one can still approximate $\VXin \psi_i(\br) \approx \int \VXACE[\{\psiini,g_i\}](\br,\br')\psi_i(\br')\dr'$. This approximation is exact when $\psi_i$'s and $g_i$'s span the same space. Thus, as long as the $\psi_i$'s do not deviate much from $g_i$'s, ACE offers a good approximation for the action of $\VXin$ on $\psi_i$'s. If and when the deviation between $g_i$'s and $\psi_i$'s exceed a tolerance, we update the ACE operator to $\VXACE[\psiini, \psi_i]$.  We note that the most expensive part of constructing the ACE operator is the evaluation of the $u_i(\br)$. To elaborate, $u_j(\br) = -\sum_i \psiini(\br) v_{i,j}(\br)$, where  $v_{i,j}(\br) = \int \frac{\psiini(\br')g_j(\br')}{|\br-\br'|}dr'$ denotes a nonlocal evaluation. For efficient evaluation of $v_{i,j}$, we recast it into a Poisson problem:$-4\pi\nabla^2 v_{i,j}(\br) = \psiini(\br)g_j(\br)$. We solve it by discretizing the $v_{i,j}(\br)$ with a finite-element (FE) basis. The finite-element approach reformulates these calculations as a series of small, dense matrix multiplications — an operation that is exceptionally well-suited to modern computer hardware, since it performs many arithmetic operations per unit of data moved. This efficiency also makes the method naturally compatible with graphics processing units (GPUs), which substantially accelerate the computations further~\cite{das2022dft,das2023large}.

\subsection*{Ab initio densities}
The exact groundstate densities are obtained from heat bath CI (HBCI)~\cite{Holmes2016, holmes2017, sharma2017, Li2018, Dang2022, Chien2018}, a systematically convergent route to the full CI (FCI) limit with significantly reduced computational cost~\cite{Holmes2016, sharma2017, Dang2023}. In HBCI, the variational space is constructed iteratively: starting from a reference state, new determinants are selectively added to expand the wavefunction toward the FCI limit, retaining only those whose coupling to the current wavefunction exceeds a selection threshold $\varepsilon_1$. In this work, we use $\varepsilon_1 = 5 \times 10^{-5} $Ha. All electrons were correlated. Gaussian basis sets were used for all systems except $\text{C}_2 \text{H}_4$, and the basis sets were obtained from the Basis Set Exchange ~\cite{pritchard2019new}. Specifically, the cc-pV5Z basis was used for LiH ~\cite{prascher2011gaussian, kendall1992electron}, aug-cc-pVQZ for CO ~\cite{kendall1992electron} and cc-pVTZ for Ne and benzyne~\cite{dunning1989gaussian}. For $\text{C}_2 \text{H}_4$, the HBCI calculation was performed using the TZ2P Slater-type basis from the ADF program~\cite{Van2003}. The density for ortho-benzyne (C$_6$H$_4$) was taken from Ref.~\cite{kanungo2019exact}; details of the corresponding wavefunction calculations are provided therein. 

\section*{Data availability}
The molecular geometries, CI density matrices, and corresponding AO basis sets used in this study are provided in the Supplementary Information. The three-dimensional exchange-correlation potential field data generated in this study are available from the corresponding author upon reasonable request.

\section*{Code availability}
The inverse DFT code (\invDFT) is publicly available at https://github.com/dftfeDevelopers/invDFT. The inverse GKS calculations presented in this work were performed using a private development branch of \invDFT, which will be made available upon reasonable request to the corresponding author.

\section*{Acknowledgment}
We acknowledge the support of Department of Energy, Office of Science, through grant number DE-SC0022241, under the auspices of which this study is conducted. This study used resources of the Oak Ridge Leadership Computing Facility, which is a DOE Office of Science User Facility supported under Contract DE-AC05-00OR22725 . This study also used resources of the NERSC Center, a DOE Office of Science User Facility using NERSC award BES-ERCAP0034270.

\bibliographystyle{unsrt}
\bibliography{ref}

\end{document}